\documentclass[twocolumn]{aastex701}

\newcommand{\pks}{PKS~1830$-$211}
\newcommand{\g}{$\gamma$}
\newcommand{\Fermi}{$Fermi$}
\newcommand{\tab}{$\tau_{AB}$}
\newcommand{\magRatioG}{$\mu_\gamma$}

\newcommand{\averageRadio}{$25.0 \pm 2.3$}
\newcommand{\paperI}{\textsc{Paper~I}}
\newcommand{\paperII}{\textsc{Paper~II}}
\newcommand{\paperIII}{\textsc{Paper~III}}

\graphicspath{{./in_use_figures/}}

\usepackage{caption,subcaption}
\usepackage{hyperref}
\usepackage{booktabs}

\usepackage[export]{adjustbox}
\usepackage{tabularx}

\received{\today}
\revised{\today}

\shorttitle{Macrolensing effects in \pks}
\shortauthors{Buson et al.}

\begin{document}

\title{Gamma-ray Time Delay and Magnification Ratio in the Gravitationally-Lensed Blazar \pks}

\author[orcid=0000-0002-3308-324X,sname='Sara Buson']{S. Buson}
\affiliation{
       Julius-Maximilians-Universit\"at W\"urzburg, Fakultät f\"ur Physik und Astronomie, Institut f\"ur Theoretische Physik und Astrophysik, Lehrstuhl f\"ur Astronomie,  Emil-Fischer-Str. 31, D-97074 W\"urzburg, Germany}
\affiliation{
       Deutsches Elektronen-Synchrotron DESY, Platanenallee 6, 15738 Zeuthen, Germany}
\email[show]{sara.buson@desy.de}  

\author[]{M. De Toma} 
\affiliation{SISSA, via Bonomea 265, I-34135 Trieste, Italy}
\email[show]{mdetoma@sissa.it}

\author[]{S. Larsson} 
\affiliation{Department of Physics, KTH Royal Institute of Technology, AlbaNova, SE-106 91 Stockholm, Sweden}
\affiliation{The Oskar Klein Centre for Cosmoparticle Physics, AlbaNova, SE-106 91 Stockholm, Sweden}
\email[show]{stefan@astro.su.se}

\author[]{C. C. Cheung} 
\affiliation{Space Science Division, Naval Research Laboratory, Washington, DC 20375, USA}
\email[]{chi.c.cheung2.civ@us.navy.mil}

\author[]{P. Cristarella Orestano} 
\affiliation{Istituto Nazionale di Fisica Nucleare, Sezione di Perugia, I-06123 Perugia, Italy}
\affiliation{Dipartimento di Fisica, Università degli Studi di Perugia, I-06123 Perugia, Italy}
\email[]{paolo.cristarellaorestano@pg.infn.it}

\author[]{S. Ciprini} 
\affiliation{Istituto Nazionale di Fisica Nucleare, Sezione di Roma “Tor Vergata,” I-00133 Roma, Italy}
\affiliation{Space Science Data Center - Agenzia Spaziale Italiana, Via del Politecnico, snc, I-00133, Roma, Italy}
\email[]{stefano.ciprini.asdc@gmail.com}

\author[]{M. Razzano} 
\affiliation{Department of Physics, University of Pisa, Largo B. Pontecorvo, 3, 56127, Pisa, Italy}
\affiliation{Istituto Nazionale di Fisica Nucleare, Sezione di Pisa, Largo B. Pontecorvo, 3, 56127, Pisa, Italy}
\email[]{massimiliano.razzano@unipi.it}

\author[]{A. Dominguez} 
\affiliation{IPARCOS and Department of EMFTEL, Universidad Complutense de Madrid, E-28040 Madrid, Spain}
\email[]{alberto.d@ucm.es}

\author[]{M. Ajello} 
\affiliation{Department of Physics and Astronomy, Clemson University, Kinard Lab of Physics, Clemson, SC 29634-0978, USA}
\email[]{majello@clemson.edu}

\author[]{S. Cutini} 
\affiliation{Istituto Nazionale di Fisica Nucleare, Sezione di Perugia, I-06123 Perugia, Italy}
\email[]{sara.cutini@pg.infn.it}



\begin{abstract}
\noindent
We present the characterization of macrolensing properties of the gravitationally lensed system \pks, utilizing data from the \Fermi\ Large Area Telescope.
While at \g-rays we can not spatially resolve the lensed images, a macrolensing-induced time pattern is expected in the blazar’s lightcurve, resulting from the delay between variable \g-ray components originating from its two brightest lensed images. Compared to our previous study, here we employ high-quality lightcurves coupled with prolonged outburst activity, and improved time-series techniques. Analyzing six independent data segments,  we identified a delay of $20.26 \pm 0.62$ days (statistical and stochastic uncertainty), with a chance detection probability at the $2.5 \times 10^{-5}$ level (post-trial). We also present a novel approach to the magnification ratio estimate based on a comparison with simulated data. Our work suggests that the \g-ray flux ratio between the two main lens components is \magRatioG~$\lesssim 1.8$. We do not observe convincing evidence of microlensing effects, as previously claimed.
The measured \g-ray time delay is in $2\sigma$ tension with radio-based estimates, suggesting either distinct emission sites 
or \g-ray production in a region opaque to radio.
Our study highlights the potential of well sampled lightcurves and advanced time‑series techniques to distinguish true lensing-induced delays from stochastic variability.
When combined with improved 
lens models, \pks\ and similar sources constitute promising systems for time-delay cosmography, offering new insights into both jet structure and cosmological parameters.

\end{abstract}



\section{Introduction}
\label{sec:intro}

Gravitational lensing presents a distinctive opportunity for probing mass distributions across the Universe and measuring the Hubble constant ($H_0$). Recent studies 
have demonstrated the efficacy of this approach in cosmological investigations using the time-delay cosmography method (\citealp{Refsdal:1964}; see also findings from the H0LiCOW collaboration \citealp{H0LiCOW:2016tzl}). At the same time, they highlighted the importance of accurate time-delay measurements in lensed systems.

The system \pks\ is a notable example of strong gravitational lensing.
Initially catalogued as a bright radio source at 5~GHz in the Parkes survey \citep{Lindman99}, later observations by the Very Large Array (VLA) and Australian Telescope Compact Array (ATCA) revealed a lensed system composed of two compact highly magnified components (hereafter images A and B, to the northeast and southwest, respectively) separated by $\sim 1\arcsec$ and surrounded by an Einstein ring \citep{PrameshRao88,Jauncey91}. The lensed background object is a flat-spectrum-radio-quasar (FSRQ) at redshift z=2.507. Evidence for an additional fainter component has been tentatively discussed \citep{Subrahmanyan:1990,Nair:1993} 
with the most recent ALMA observations pinpointing a point-like radio source, $\sim0.3\arcsec$ away from image B,
attributed as the central, third lensed image \citep{Muller:2020}.

The system is a compound gravitational lens, with two distinct lensing galaxies involved \citep{Lovell96}. The main lensing galaxy is a foreground face-on spiral galaxy $z = 0.89$ that is also responsible for molecular absorption features \citep{Wiklind96}. The detection of additional HI and OH absorption revealed a second intervening galaxy at $z = 0.19$ \citep{Lovell96}, which should have a limited influence on the lensing effects. Studies suggest that its contribution to the gravitational lensing phenomenon is relatively minor, mainly due to its lower mass and distance compared to the main lensing galaxy at $z=0.89$. Consequently, its impact on the overall lensing dynamics and image distortion of \pks\ is considered negligible \citep[e.g.,][]{Winn:2002}.

Despite the rich set of observations collected for \pks\ at several wavelengths (radio, optical, X-rays and \g-rays), the measurement of the temporal delay between the A/B components (hereafter \tab) remains uncertain, with several estimates debated in the literature
\citep{vanOmmen:1995,Lovell98,Wiklind:2001,Barnacka11,Neronov15,Barnacka:2015,Abhir:2021,Muller:2023,Agarwal:2025,Wagner:2025,Biggs:2025}. In our previous study \citep[][hereafter \paperI]{PKS1830_LAT:2015}, we reported no convincing evidence for lensing effects in the \g-ray lightcurve between $2008-2011$. 
Following a prolonged \g-ray high state in 2019–2020 \citep{Buson_PKS1830_LAT_Atel:2019}, we initiated a multiwavelength campaign including X-ray and radio observations \citep{Buson_XRT_ATel:2019}. The results of our  follow-up monitoring at 43 GHz with the Very Long Baseline Array (VLBA)  follow-up are presented in \citet[][\paperIII]{spingola:2025}.
In this paper (\paperII), using over a decade-long legacy $Fermi$-LAT observations and advanced time-series techniques, we resolve the long-debated \g-ray lensing properties of this system, and unambiguously measure the macrolensing time delay \tab\ along with limits in the magnification ratio \citep[early findings appeared in][]{deToma_Master:2023,DeToma_ICRC:2025}.

The paper is organized as follows. Sec.~\ref{sec:time_delay} offers an overview of the current state of the art. In Sec.~\ref{sec:LATdata} describes the \textit{Fermi}-LAT observations and data analysis. Sec.~\ref{sec:time_series} introduces time-series methods for estimating the time delay and presents a novel approach to compute the macrolensing magnification ratio.
Sec.~\ref{sec:method} outlines the strategy for assessing the local and global statistical significance of the findings. In Sec.~\ref{sec:results} we present the results. Our new findings are discussed in the context of previous discrepancies in delay estimates in Sec.~\ref{sec:comparison}, and of claimed microlensing effects and variability of the magnification ratio in Sec.~\ref{sec:micro}. Sec.~\ref{Sec:conclusions} concludes summarizing the main results.

\begin{figure*}[ht!]
    \centering
\includegraphics[width=.95\linewidth]{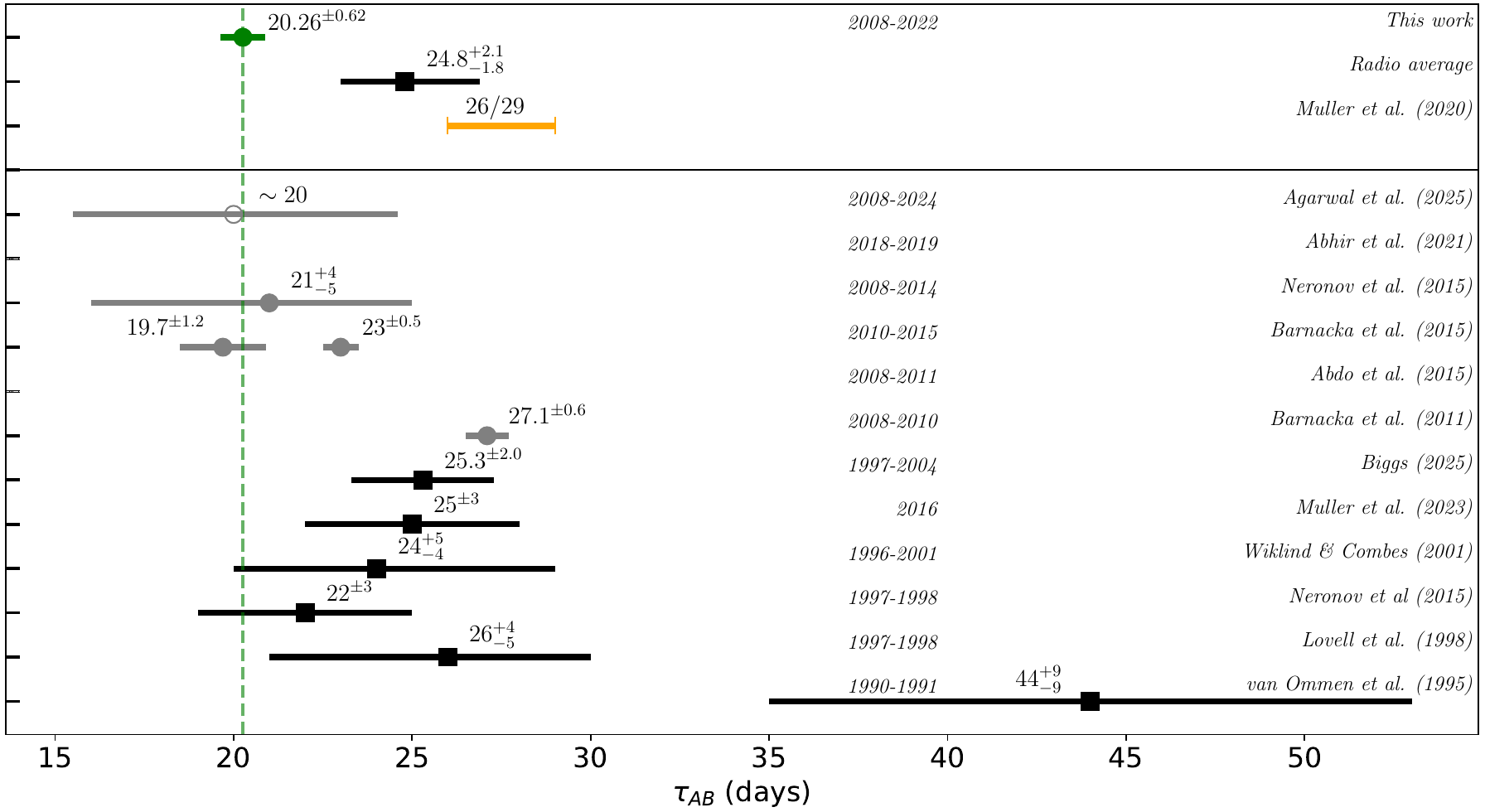}
    \caption{
    Estimates of the time delay \tab\ between the A and B images of \pks. The value measured in this work (green) is in mild tension with the radio-average estimate from the literature (black), and with predictions from lens modeling based on radio observations (yellow).
    The lower panel reports \g-ray estimates (gray, filled circles), and radio estimates (black, filled squares).
    The study of \citet{Agarwal:2025} reports an approximate delay estimate; in \paperI\ and in \citet{Abhir:2021}, no delay measurement was inferred.
    The corresponding observational time windows and references are reported in italics.
    }
    \label{fig:literature_delay}
\end{figure*}
%
%
 \section{Time Delay and Magnification ratio}
 \label{sec:time_delay}
Gravitational lensing is an achromatic effect, hence there is no dependency on the observing wavelength. Therefore, if the overall electromagnetic emission of the blazar originates from a common physical region, and  there are no extrinsic propagation effect (e.g. any effect due to smaller lens structures, such as microlensing, is negligible), the observed time delay(s) between the lensed images will be the same at all bands.
%
The lively literature debate about \tab\ in the \pks\ system highlights the complexity of this system. Several methodologies have been employed to infer the time delay associated with the macrolensing effect.

\begin{itemize}
    \item 
    Delay estimates based on radio observations. 
    High-resolution instruments can spatially separate the A and B images, allowing direct measurement of their fluxes. Time-series, dispersion, $\chi^2$ minimization, or regression methods are applied to the resolved light curves.
    However, radio observations generally encompass a limited time span ($<1$ year), with few visits of irregular cadence.
    \item 
       Delay measurements based on \g-ray time series analysis. Although currently operating \g-ray instruments lack the arcsecond-spatial resolution to resolve the individual lensed components, the strongly variable emission from the blazar, combined with regular long-term flux monitoring, can reveal lensing effects. 
       Light rays from different lensed images reach the observer with relative time delays, imprinting characteristic features in the lightcurves that can be used to study gravitational lensing and related physical properties.
    \item 
    Predictions inferred from the modeling of the lens system. A major challenge comes from the location of \pks, which lies near the Galactic plane. The vicinity to the galactic center (l$=12.2^\circ$, b$=-5.7^\circ$), makes optical identification difficult as optical observations are affected by dust extinction and by contamination by the Milky Way. The most refined lens model is based on radio observations \citep{Muller:2020}.
\end{itemize}

Figure \ref{fig:literature_delay} summarizes the estimates of \tab\  based on radio observations (black), \g-ray time series (gray), and lens modeling (yellow). For completeness, the upper row reports the value derived from this study (green;  \ref{subsec:delay}). Radio-based values converge toward an average\footnote{The estimate from \citet{vanOmmen:1995} was excluded, as likely affected by contamination from the Einstein ring. Similarly, we excluded the result from \citet{Neronov15} and \citet{Lovell98}, adopting instead the latest analysis by \citet{Biggs:2025} based on the same data.} \tab~= \averageRadio\ days. Estimates at \g-rays were controversial, with multiple \tab\ values proposed, not always consistent with each other.
In some cases, no significant evidence for a trailing component was found \citep[\paperI,][]{Abhir:2021}. 

The magnification ratio, defined here as the ratio of the fluxes of the leading and trailing components
$\mu = F_A(t) / F_B(t+\tau_{AB}) $, ranges between $1.0 - 2.0$ according to radio-based estimates \citep{vanOmmen:1995,Frye_1997,Lovell98,Wiklind_1998,Muller:2020, Muller:2023}. Previous measurements carried out at \g-rays appeared in tensions with this value, with \magRatioG\ ranging between $\sim2-7$ and invoked also microlensing effects \citep{Neronov15,Barnacka:2015,Agarwal:2025}.

\section{{\it Fermi}-LAT OBSERVATIONS AND ANALYSIS} \label{sec:LATdata}

The $Fermi$-LAT data are reduced with the Python package \texttt{fermipy} \citep[
v1.2.0,][]{Wood:2017}, using the \texttt{P8R3\_SOURCE\_V3} instrument response functions \citep[IRFs][]{Atwood:2009,Bruel:2018}. We select photons of the \texttt{Pass 8 SOURCE} class, in a region of interest (ROI) of 10$^\circ$ $\times$ 10$^\circ$ square, centered at the target, within the time span 2008-08-04 to 2023-08-22 (MJD $54682 - 60178$). To minimize the contamination from $\gamma$ rays produced in the Earth’s upper atmosphere, a zenith angle cut of $\theta < 90^{\circ}$ is applied, along with standard data quality cuts ($ \rm DATA\_QUAL > 0) \&\& (LAT\_CONFIG == 1$). Due to the vicinity of the target to the galactic center, a lower energy cut of $\geq 0.2$~GeV is applied to mitigate the impact of the bright emission from the galactic components.
The ROI model includes all \Fermi-LAT Fourth Source Catalog  Data Release 3 sources \citep[4FGL-DR3,][]{4FGL_DR3:2022} located within 15$^{\circ}$ from the ROI center, as well as the Galactic and isotropic diffuse emission \footnote{\href{https://fermi.gsfc.nasa.gov/ssc/data/access/lat/BackgroundModels.html}{LAT Background Models}} (\texttt{gll\_iem\_v07.fits} and \texttt{iso\_P8R3\_SOURCE\_V2\_v1.txt}). 

A binned maximum likelihood  analysis in the $0.2-300$ GeV energy range is performed, following \citep[][]{Penil_trends:2025}.
In the analysis, the test-statistics (TS) is defined as $2\log(L/L_0)$, where \textit{$L$} is the likelihood of the model with a point source at a given position and \textit{$L_0$} is the likelihood without the source \citep{mattox1996}.
\begin{figure*}[t!]
    \centering
\includegraphics[width=0.95\linewidth]{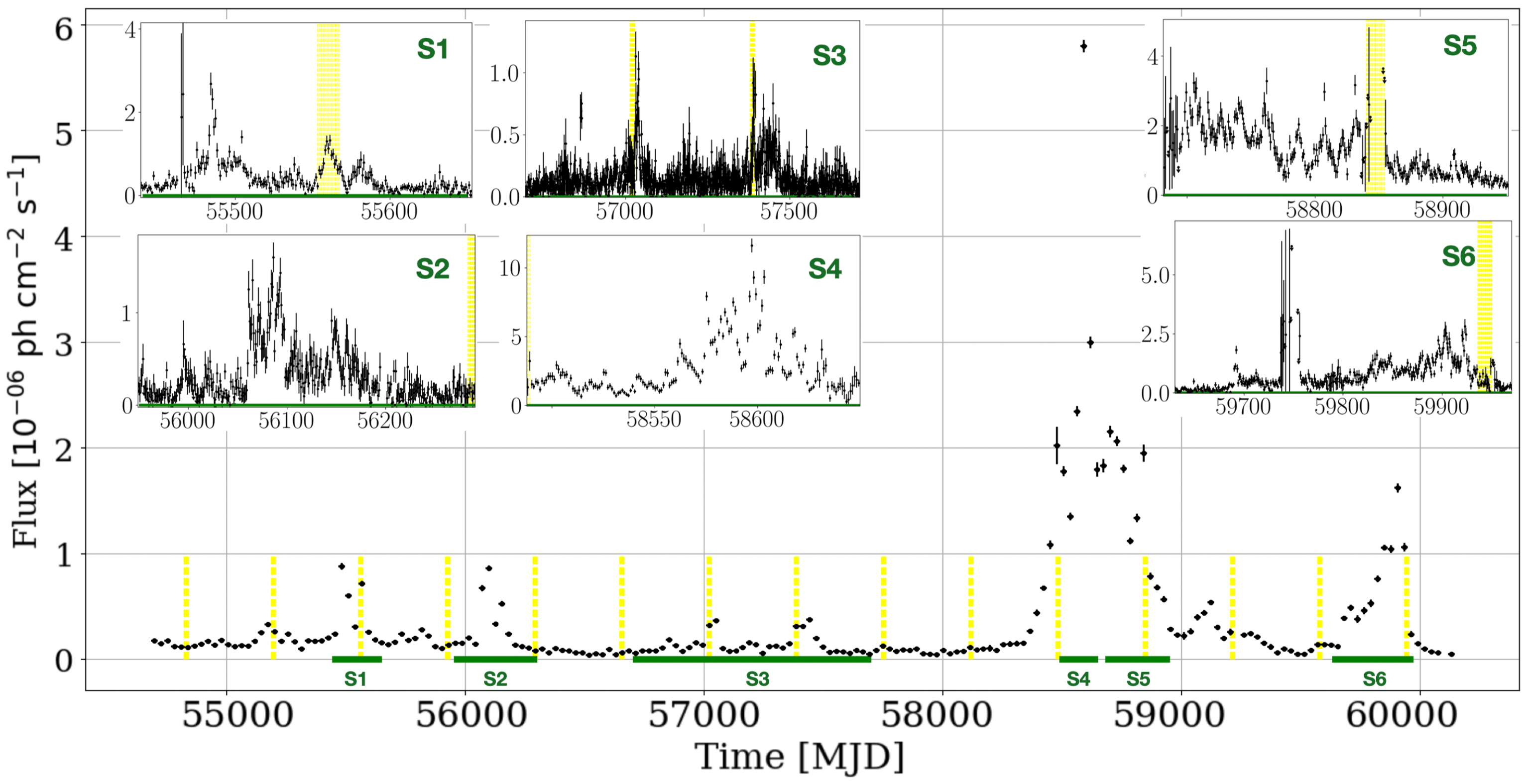}
    \caption{Strong variability is evident throughout the entire 28-day bin LAT LC of \pks\ shown in the main panel. The six active states utilized in the time-series analysis (green) are listed in Table \ref{tab:segments}. A close-up view of these segments is presented in the insets, using a 1-day binning. Dashed yellow lines indicate the time intervals when the Sun is within $8^\circ$ of the ROI center, which were censored in the time-series analysis.  Down arrows represent 95\% upper limits. 
    }
    \label{fig:LAT_LC}
\end{figure*}
%
The LC for the time-series analysis is produced with 1-day binning; 28-day bin and adaptive binning LCs are built for illustration (Section \ref{sec:crabToO}). 
We consider the target source to be detected when TS$>$1 in the corresponding time bin; otherwise,  a 95\% confidence upper limit is reported. 

Given the target’s proximity to both the ecliptic and the Galactic plane, we apply additional LC filtering to optimize the number of flux points with better signal-to-noise ratio information. To mitigate contamination by quiet Sun emission, we discard time intervals when the Sun is within $8^{\circ}$ of the ROI center \citep{Quiet_Sun:2011}.
In the time-series analysis, we retain only points with relative error less than 100\% and  with TS$>1$.

\section{Time-Series Analysis} \label{sec:time_series}

\subsection{The auto-correlation function}

The auto correlation function (ACF) is a widely used method to quantify the similarity or correlation between a time series and a time shifted copy of itself, i.e. the ACF is a function of this time shift or “lag”. We adopt the “local” discrete ACF ($\text{DACF}_L$), an alternative definition of the ACF, which has been shown to be more effective compared to the standard DACF definition in measuring time lags in AGN LCs \citep{Welsh99}. The $\text{DACF}_L$ includes only those values of the time series $a_i$ that overlap at a given lag, $\tau_m = m \Delta t$, to determine the mean and standard deviations. It is defined as:
\begin{equation}\label{Eq:local_DACF}
    \text{DACF}_{L}(\tau_m) = \frac{\frac{1}{N-m} \sum_{i=1}^{N-m}(a_i-\bar{a}_1)(a_{i+m}-\bar{a}_2)}{d_1 d_2}
\end{equation}
where
\begin{equation}
    d_1 = \left[\frac{1}{N-m}\sum_{i=1}^{N-m}(a_i-\bar{a}_1)^2\right]^{1/2}
\end{equation}
\begin{equation}
    d_2 = \left[\frac{1}{N-m}\sum_{i=m+1}^{N}(a_i-\bar{a}_2)^2\right]^{1/2}
\end{equation}
\begin{equation}
    \bar{a}_1=\frac{1}{N-m}\sum_{i=1}^{N-m}a_i, \hspace{1cm} \bar{a}_2=\frac{1}{N-m}\sum_{i=m+1}^N a_i,
\end{equation}
where the time series $a_i$ is sampled at discrete times $t_i$ ($i=1,...,N$) with equal sampling ($\Delta t= t_{i+1}-t_i$).

\subsection{Macro trends and detrending method}\label{subsec:detrending}
Often blazars' LCs are characterized by a dominant low-frequency power due to a red-noise component, along with the high-frequency fluctuations. The DACF tends to exhibit a smooth and long-tailed decay, reflecting the stochastic nature and long-timescale variability of the blazar's emission. Overall, the presence of strong red-noise variability can smooth or wash out any sharp features in the DACF that might arise from periodic or quasi-periodic signals. 
A viable strategy to minimize the impact of the macro-trends consists in applying a detrending that “pre-whitens” the data,
as investigated in \citet{Welsh99}. We develop a new detrending technique based on a spline function that subtracts slow variability using a cubic piecewise polynomial. Simulation tests indicate that a 15-day spline bin provides the best performance for this case \citep{deToma_Master:2023}.

\subsection{Strategy for choice of time intervals}\label{Subsec:time_segments} 
When searching for periodic or quasi-periodic patterns in a  LC, it is common to analyze a single, long or highly variable segment. While this approach can seem rigorous, in the presence of strong variability and stochastic processes, the true global significance of any detected pattern is hard to quantify \citep{Vaughan:2016}.

Red‑noise dominated LCs can easily produce few‑cycle modulations that appear locally significant, and the probability of detecting such false positives increases when a large ensemble of such sources is available (Appendix \ref{app:checks}). Unless there are well-motivated priors favoring specific sources or time scales, such approach can lead meaningful findings only when coupled with a rigorous way to account for the look-elsewhere effect \citep{Penil:2020,Penil:2024,Adhikari:2024,Penil_mwl_variability:2024,Penil_trends:2025,Rico:2025,Penil_distortions:2025,Penil_kink:2025,Adhikari_optical:2025,Penil:2025}.

In this study, we adopt a more balanced strategy, applying the time-series analyses to several distinct, non-overlapping LC-data segments which qualitatively display well-defined time-variable structures. Our strategy is not only more robust and conservative but also offers other advantages. On the one hand, we can test possible variations of the time patterns over time; on the other hand, observing the same one consistently across multiple segments corroborates its authenticity.

The number of data points required in a time-series analysis depends on several factors. \citet{Welsh99} suggests that the LC length should be at least $\sim 10$ times the lag of interest. For \pks, prior studies indicate a timescale \tab\ of $\sim 20$–30 days. Consequently, we select segments that span $\sim 200$ days, and explore lags between 10 to 80 days. Simulations demonstrate that further increasing the length of a LC dominated by red noise may not lead to a significant improvement in the time pattern measurement \citep{Welsh99}. The reason is that the DACF is primarily influenced by the segments of the LC with the strongest variability, while the contribution from comparatively less variable parts becomes minimal or negligible. In particular, when the uncertainty of the correlation measure is dominated by stochastic noise, segmenting the data for analysis is advantageous. 
Following these prescriptions, we select the six most active data segments which are listed in Table~\ref{tab:segments} and highlighted as the insets in Figure~\ref{fig:LAT_LC}.

\section{Methodology for statistical significance and time delay estimate}
\label{sec:method}
\subsection{Local significance of the DACF peaks}
\label{subsec:local_sign_method}
To estimate the significance of time lags in a given LC segment, we generate $10^4$ mock LCs using the method by \citet{Emmanoulopoulos13}, described in \citet{Cristarella:2025}\footnote{Available at: \href{https://github.com/PaoloCO42/Emmanoulopoulos-Light-Curve-Simulations}{https://github.com/PaoloCO42/Emmanoulopoulos-Light-Curve-Simulations}}. These simulated LCs match the power spectral density (modeled as a power law, PSD~$\sim 1/f^\alpha$) and probability distribution function of the given segment. To account for uneven sampling due to data gaps, we apply the real data's time sampling to the mocks. Then, we compute the DACFs of these mock LCs.
For each lag $\tau$, we compare the DACF($\tau$) value found for \pks\ with the distribution of mock-DACF($\tau$) values. 
For every tested $\tau$, the DACF($\tau$) distribution is fitted with a Gaussian function, 
to infer the likelihood of observing a value similar to that of \pks\ by chance. The performance of the methodology were tested on simulated data \citep[see][]{deToma_Master:2023}, where we injected a known delayed signal into mock LCs (after detrending) and added flux-dependent white noise  with standard deviation equal to the error estimates of the observational data. The method recovered the injected delay with significance $>5\sigma$ in more than 80\% cases, with no spurious delays observed persistently above $2\sigma$.
In the following, we define a DACF peak as a lag that shows local significance $>2 \sigma$.  Unless otherwise stated, the DACFs are computed using a 1-day time bin.

\subsection{Global significance of the DACF peaks}\label{subsec:global_sign}

Random fluctuations in stochastic variability can as well produce significant DACF peaks at any lag (Appendix \ref{app:checks}). These chance probabilities can be estimated by Monte Carlo simulations where the same number of peaks as identified in the real-data DACFs are randomly distributed between lag 10 and 80. Assuming a peak lag uncertainty, or width $\Delta \tau $, there will be $70/\Delta\tau$ possible positions in the DACFs. We require all 6 DACFs to show a peak in the same bin, as a delayed signal arising from macro-lensing effect is expected to produce a peak at the same lag in all segments. Although choices of lag range or peak threshold introduce some arbitrariness, this approach avoids assumptions about the LC variability itself. 

\subsection{DACF peak properties}
\label{subsec:peak_properties}
The time lag properties of significant DACF peaks are derived by fitting a Gaussian function to the real-data DACF peak, obtaining the time lag estimate, peak height (i.e. amplitude), and width. The same analysis is then applied to the simulated LCs and the standard deviation of the mock-DACF fitted parameters is used to estimate the uncertainties for the corresponding \pks\ results, taking both statistical and stochastic variability into account.

\section{Results}
\label{sec:results}
\subsection{Long-term LC and DACF}
\begin{figure*}[ht!]
    \centering
    \includegraphics[width=0.9\linewidth]{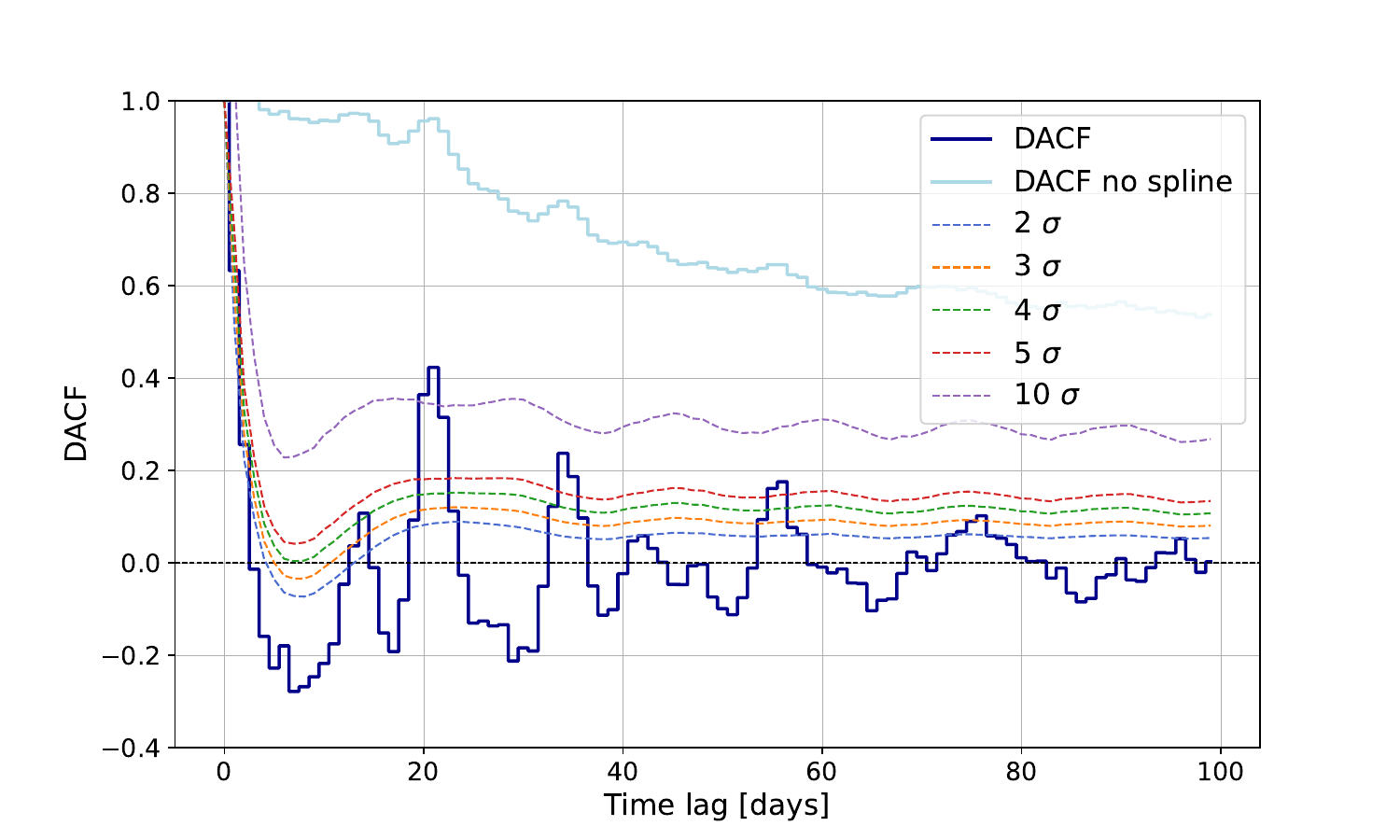}
    \includegraphics[width=\linewidth]{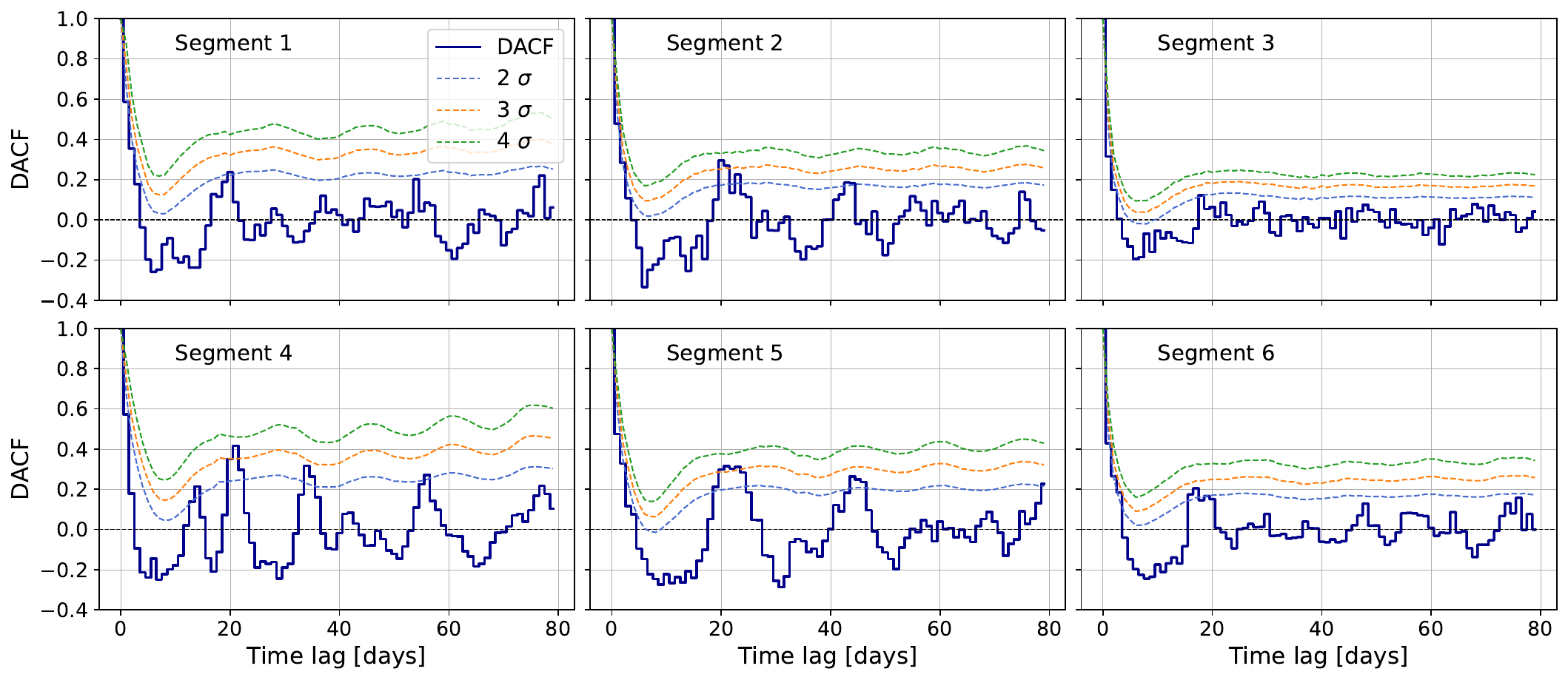}
    \caption{
    Top: DACF of the long-term 1-day bin LAT LC (MJD 54682-60178) of \pks\ is shown in cyan. The DACF obtained by the detrended LC is shown in blue, along with the local significance levels. 
    Bottom: DACF for the six time segments analyzed in this work, listed in Table \ref{tab:segments}. The respective LCs are shown in the panels of Figure~\ref{fig:LAT_LC}. The blue, orange, green dashed lines represent the $2\sigma$, $3\sigma$, $4\sigma$ local significance
(Section \ref{subsec:local_sign_method}).}
 \label{fig:segments_DACF}
\end{figure*}

The long-term 1-day binned LC, spanning MJD~54682 to MJD~60178, is shown in Figure~\ref{fig:LAT_LC}, with the corresponding DACF displayed in Figure~\ref{fig:segments_DACF}. These are included for illustration purposes only and not used in estimating gravitational lensing effects, except for consistency checks. Although in this DACF several peaks exceed a local statistical significance of $5\sigma$, with one at $\sim20$ days reaching $>10\sigma$, their global significance is difficult to assess, as the contribution from stochastic variability is hard to disentangle. This is because the long-term LC includes both quiescent and highly active phases, making it overly simplistic to model the entire period with a single PSD. We address this issue by examining the LCs of two similarly bright blazars in Appendix~\ref{app:checks}.
Moreover, the large flux variability across the LC results in the overall DACF being dominated by the brightest epoch, i.e. the flaring episode around MJD~58600.


\subsection{Segments DACFs results} \label{subsec:segments_results}
DACFs of the six selected data segments are shown in Figure~\ref{fig:segments_DACF} along with the statistical significance thresholds obtained as described in Section~\ref{subsec:local_sign_method}. The fact that the DACF of segment 4 in Figure~\ref{fig:segments_DACF} is almost identical to the DACF of the long-term LC confirms that the latter is dominated by the strong activity in that segment.

\begin{figure*}[ht!]
  \centering
  \begin{minipage}{0.5\textwidth}
    \flushleft
    \begin{tabular}{lccccccc}
    \toprule
    & Start & Stop & Points &PSD & Time lag  & Sign. \\
    & [MJD] & [MJD]&    & index  & [days]   &  [$\sigma$] \\
    \midrule
    1 & 55440 & 55650 & 170 & $-1.24 \pm 0.09$ & $19.58 \pm 1.12$ & $\geq2$ \\
    2 & 55950 & 56300 & 265 & $-0.93 \pm 0.09$ & $21.49 \pm 1.06$ & $\geq3$ \\
    3 & 56700 & 57700 & 655 & $-0.84 \pm 0.05$ & $20.22 \pm 1.23$ & $\geq2$ \\
    4 & 58487 & 58650 & 159 & $-1.22 \pm 0.13$ & $20.96 \pm 0.75$ & $\geq3$ \\
    5 & 58680 & 58950 & 246 & $-1.16 \pm 0.09$ & $22.03 \pm 0.83$ & $\geq3$ \\
    6 & 59630 & 59970 & 271 & $-0.86 \pm 0.08$ & $18.72 \pm 0.86$ & $\geq2$ \\
\midrule
\multicolumn{3}{c}{\textbf{1--6 combined}} & \textbf{1766} & & $\textbf{20.26} \pm \textbf{0.62}$ & \\
\bottomrule
    \end{tabular}
  \end{minipage}
  \hfill
  \begin{minipage}{0.47\textwidth}
    \centering
    \includegraphics[width=0.8\linewidth]{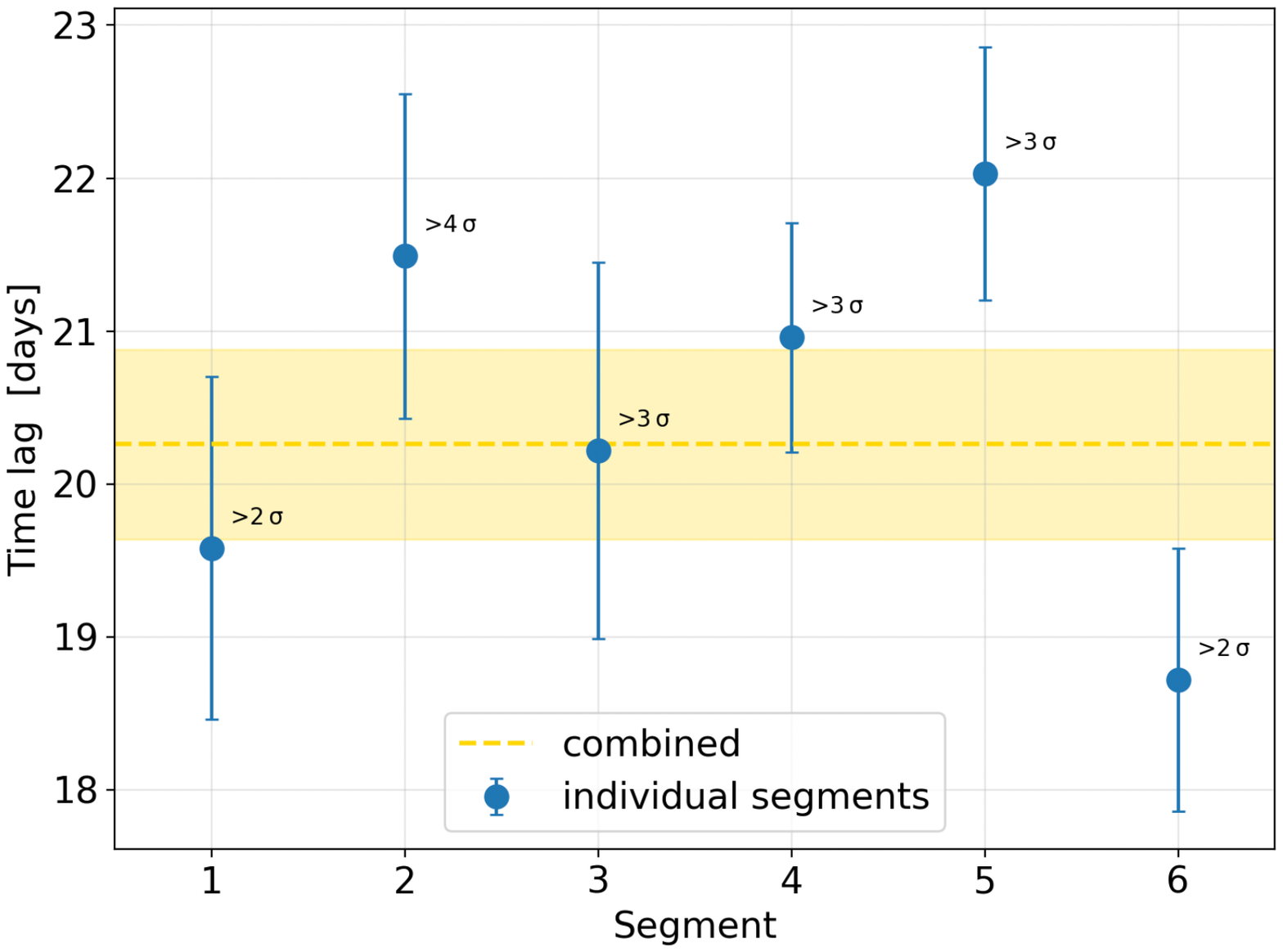}
  \end{minipage}
    \captionof{table}{Time lag estimates for the six LC segments analysed in Section~\ref{sec:time_series}, along with their local significances.
    Uncertainties are derived from simulations that reproduce the statistical and sampling properties of each segment, adopting the segment-specific PSD index (Section~\ref{sec:method}). The row labeled “$1-6$ combined” presents the DACF result for a combined LC that includes all segments, each normalized to the same mean flux, shown in Figure~\ref{fig:acfs_segm} (top). A visual summary is provided  on the right.
    }
    \label{tab:segments}
\end{figure*}

Time lags were estimated individually for each of the six data segments as described in Section \ref{subsec:peak_properties}. 
Among those with local significance above 2$\sigma$, only one, around 20 days, shows a DACF peak that is consistently present across all segments. The best-fit peak values and uncertainties for this lag are reported in Table~\ref{tab:segments},  for each segment.
Following Sec.~\ref{subsec:global_sign}, we estimate its global chance probability. The observed number of peaks in the six segment DACFs is 1, 2, 1, 4, 2, and 1, respectively. Assuming a typical peak width of $\Delta \tau \sim 5$, this yields a global chance probability of $2.5 \times 10^{-5}$. We interpret the observed time lag of $\sim$20 days as arising from macrolensing effect and, thus, as representing $\tau_{AB}$. Radio studies have indicated that the northeast component A (leading) precedes the southwest component B (trailing). However, from the \g-ray analysis we cannot unambiguously determine which component leads.
 
%

\subsection{$\tau_{AB}$ time delay }
\label{subsec:delay}
The values estimated for $\tau_{AB}$ in Table \ref{tab:segments} for the six segments are consistent with each other, within their uncertainties. 
Furthermore, one may note that even if the DACF lag estimate for the brightest segment~4 has overall lower statistical errors due to reduced measurement noise than for other segments, the stochastic uncertainty is about the same. This shows that not accounting for stochastic effects would result in an underestimation of the uncertainties.

\begin{figure}[ht!]
\caption{
Top: 
The 1-day binned DACF from the combined six data segments, each normalized to the same mean flux and detrended via the spline method (Sec.~\ref{subsec:delay}). The inset zooms in around $\sim 20$ days, showing the 1-day binned DACF (red), its Gaussian fit (black), and the 0.2-day binned DACF (black histogram).
Bottom:
DACFs for the six data segments, vertically shifted for clarity. The only persistent feature appearing in all DACFs is marked by the vertical blue dashed line at 20 days.
}
\centering
\includegraphics[width=0.88\linewidth]{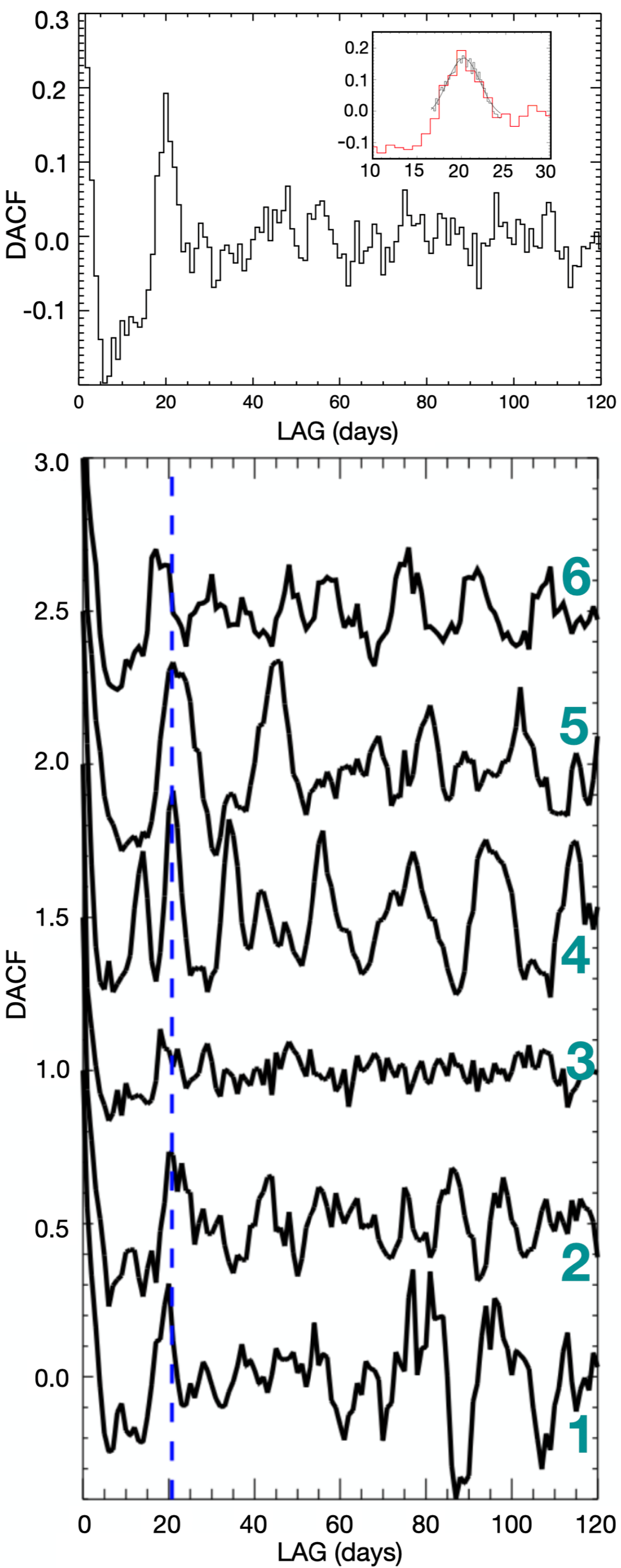}
\label{fig:acfs_segm}
\end{figure}

The impact of stochastic variations can be mitigated by combining the DACFs of the different data segments. This was done by normalizing each data segment to the same mean and then calculating a DACF for the combined LC. In this way, most of the erratic correlations get averaged out. The combined DACF and the DACFs of individual segments are shown in Figure \ref{fig:acfs_segm}. The peak around 20 days is less affected by this averaging process, strengthening its identification as a genuine, repetitive time pattern,  attributable to the presence of a trailing component produced by the gravitational lens.
The time lag for the combined LC was estimated with the same method as for the individual segments. The Gaussian fit to the correlation peak yielded a time lag of $\tau_{AB} = 20.26 \pm 0.62$ days and it is shown as an inset in Figure~\ref{fig:acfs_segm} (top). The DACF for the peak is plotted with two time resolutions, a 1-day bin (red) which is the same as the data and a 0.2-day bin (black) which is obtained by oversampling the LC segments by a factor of 10 through linear interpolation. To minimize binning effects, we use fits made to the oversampled LC, while noting that fits to the original data gave consistent results. The error estimate is again based on simulated LCs, containing stochastic red noise with a time-lag component and added Gaussian white noise to mimic measurement errors.

\subsection{Triplet effect: stochastic variability}
\label{sec:triplet}

We report the detection of characteristic structure in the DACFs, arising from the coupling of a time-lag component with rapid stochastic variability.
In the presence of a time-lag component in the LC, a DACF peak is produced at the corresponding time lag; chance correlations of rapid stochastic variability will also appear as DACF peaks at random time lags. Without any trailing component, two similar LC features, e.g. flares, will result in a single DACF peak at a time lag corresponding to the separation between the two flares. However, if both flares also have a trailing component, there will be not just one DACF peak, but three. A main central peak at the lag where both pairs of peaks overlap and two side peaks at lags where the leading and trailing components overlap. The principle of this “triplet” effect is illustrated in an idealized case in Figure \ref{fig:triplet} (top). The separation between the triplet components will be equal to the time lag of the trailing component, and their relative amplitude will depend on the lens magnification. If the magnification is 
e.g. $\mu=2$ (trailing component with half the amplitude of the leading one), the two side peaks in the DACF will be half the height of the central peak (Figure \ref{fig:triplet}, middle). For increasing magnification ratio, the relative height of the DACF side peaks will become lower. In simulations with exponential flares, the height of the side components relative to that of the central peak gradually decreased to about 0.3 at magnification ratio 2.5 and then decreased more rapidly with increasing magnification ratio.

In reality, stochastic variability is more complex and leads to a mix of overlapping features. However, by investigating the DACFs of \pks, one can pinpoint cases that appear to be produced by this triplet effect. By looking at the DACF of data segment 4 (Figure \ref{fig:triplet}, bottom),  three peaks at time lags 13.6, 34.5 and 56.1 days have lag separations (20.9 and 21.6 days) that are consistent with this effect, although the LC is more complex than the idealized case described above. 
Since the long-term LC is dominated by the brightest flare (i.e. segment 4), the corresponding DACF also shows features similar to the DACF of segment 4, including features due to the triplet effect.

A more systematic analysis of triplets in the DACF can be done by looking at the distribution of DACF peaks at long time lags. These are dominated by correlations between the different active epochs in the long-term LC. To limit the influence of other parts of the LC, we cross-correlated the six segments with each other and identified correlation peaks consistent with triplet separations. Distinct peaks were qualitatively selected and their positions and widths were estimated by Gaussian fit. An automated procedure was then used to search for triplets among the distribution of peak positions. The extracted distribution, corrected for the lag-dependent bias due to the limited segment lengths, is shown both for simulations and for \pks\ in Figure \ref{fig:mixedccfs}. The top panel shows the triplet distribution for simulations with (blue) and without (pink) a 20-day time delayed component with magnification $\mu = 1$. The corresponding distribution for \pks\ in the middle panel gives a qualitative confirmation of the presence of 20-day triplets also in the blazar DACF. Since the peak identification procedure is not fully objective, we calculated the DACFs of all the segment mixed correlations and show the average of these in the bottom panel. The peak at about 20 days is a more objective confirmation that the correlation functions contain an overabundance of features separated by $\sim 20$ days. The shallower peak around lag 40 can be understood as the higher probability of forming chance correlations between the two outer triplet components (separated by 40 days) and any randomly occurring peaks in the DACF.



\subsection{Macrolensing magnification ratio} 
\label{sec:macrolensing}

We introduce a novel method to quantify \magRatioG, exploiting the whole long-term LC information. Our approach of fitting Gaussian profiles to DACF peaks yields amplitude and width values (Section~\ref{subsec:peak_properties}), which can be used to estimate the magnification ratio \magRatioG\ by comparison with simulations. 
To model the observed DACF peak amplitude, we generated 100 simulated LCs with 0.1-day resolution, each including a delayed component shifted by 20.0, 20.1, or 20.6 days. These were then re-binned and resampled to match the 6 observed segments. Since measurement noise strongly affects the DACF peak amplitude, particular care was taken in modeling the measurement errors as a function of flux for the actual observations. Using this error model, flux-dependent Gaussian white and red noise was added to each simulated LC point. This simulation procedure was repeated for 17 magnification ratios between the trailing and leading components, ranging from 5.00 to 1.00.

\begin{figure}[ht!]
\centering
    \centering
    \includegraphics[width=0.95\linewidth]{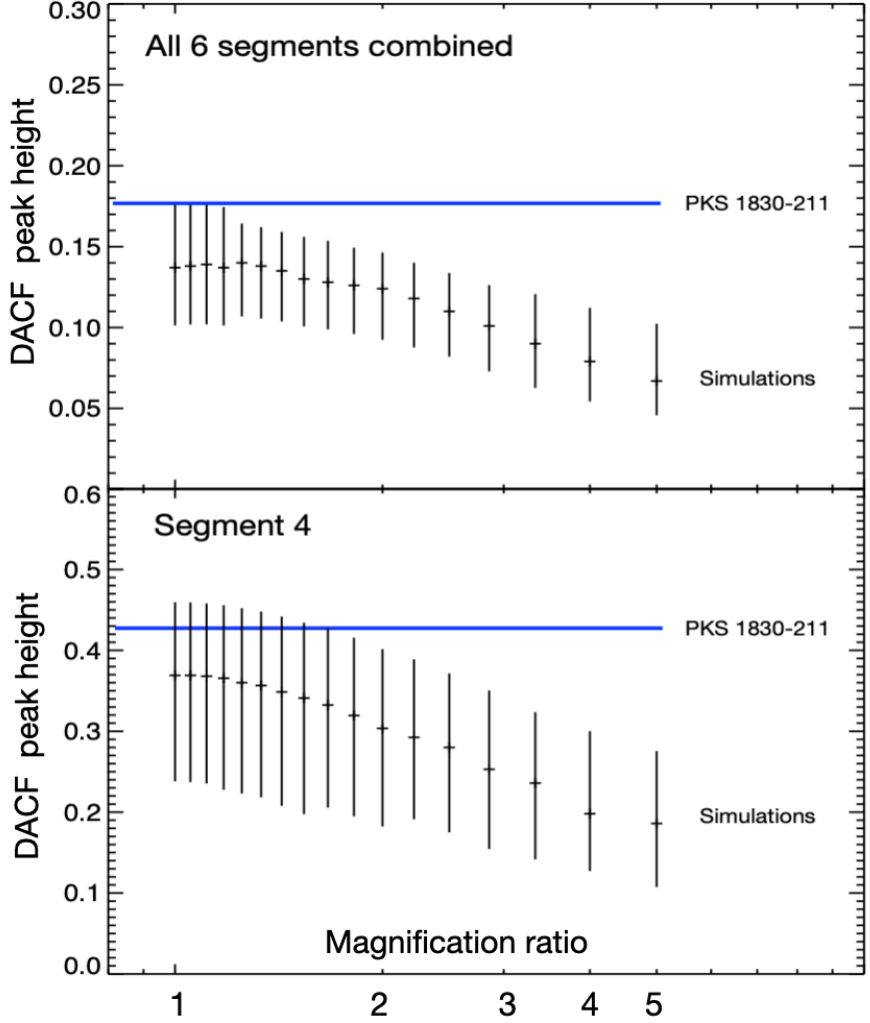}
    \caption{
    Comparison of DACF peak height for \pks\ (blue) with simulations with magnification ratios from 1.00 to 5.00. The median and the asymmetric standard deviation are shown for each simulated magnification ratio (black points). The upper panel shows the result for a LC combining all six data segments, while the lower one is based only on segment 4, where noise uncertainties are substantially lower.
    }
    \label{fig:magnificationmodel}
\end{figure}

Figure \ref{fig:magnificationmodel} shows the comparison between the peak height for \pks\ with the corresponding simulated data. The upper plot shows the results for the combined flux-normalized LC segments shown in Figure \ref{fig:acfs_segm} (top). The error bars represent the standard deviation for the DACF peaks of the 100 simulations. The wide distribution is dominated by stochastic variations between the different LC simulations. So even if the peak height and magnification ratio can be determined with a higher accuracy in a particular realization, observational data set, the uncertainty in the estimate of the intrinsic magnification ratio is substantially larger. 

At face value, Fig.~\ref{fig:magnificationmodel} suggests a magnification ratio of \magRatioG\ $\sim 1$, with the measured value at $1\sigma$ above the simulations. However, DACF peak heights are sensitive to the variance of other variability components, here, measurement noise. If this variance is overestimated, the simulated peaks will be too low. A 20\% overestimate is enough to give a shift corresponding to the difference between the simulations and blazar levels in the figure. 


Since the relative measurement errors are smaller in segment 4 than in the other segments, we have repeated the comparison using only the data and simulations for this segment. This is shown in the lower plot of Figure~\ref{fig:magnificationmodel}. Lower measurement noise yields a higher DACF peak, as visible from the higher overall DACF height values. Although some systematic uncertainty also remains for this case, these findings imply a low \magRatioG\ $\lesssim 1.8$, while comparatively larger values of \magRatioG, as those previously reported, appear unlikely. The latter may be reconciled with the specifics of previous analysis (Section \ref{sec:micro}).

An additional result from the DACF modeling of the combined 6 segments is that there is no significant difference in the width (not shown) of the DACF peak between \pks\ and the simulations. Thus, variations in lag properties reported in Table \ref{tab:segments} and visible in Figure \ref{fig:acfs_segm} are consistent with stochastic and error variability combined with data sampling.

\begin{figure*}
    \centering
    \includegraphics[width=.85\linewidth]{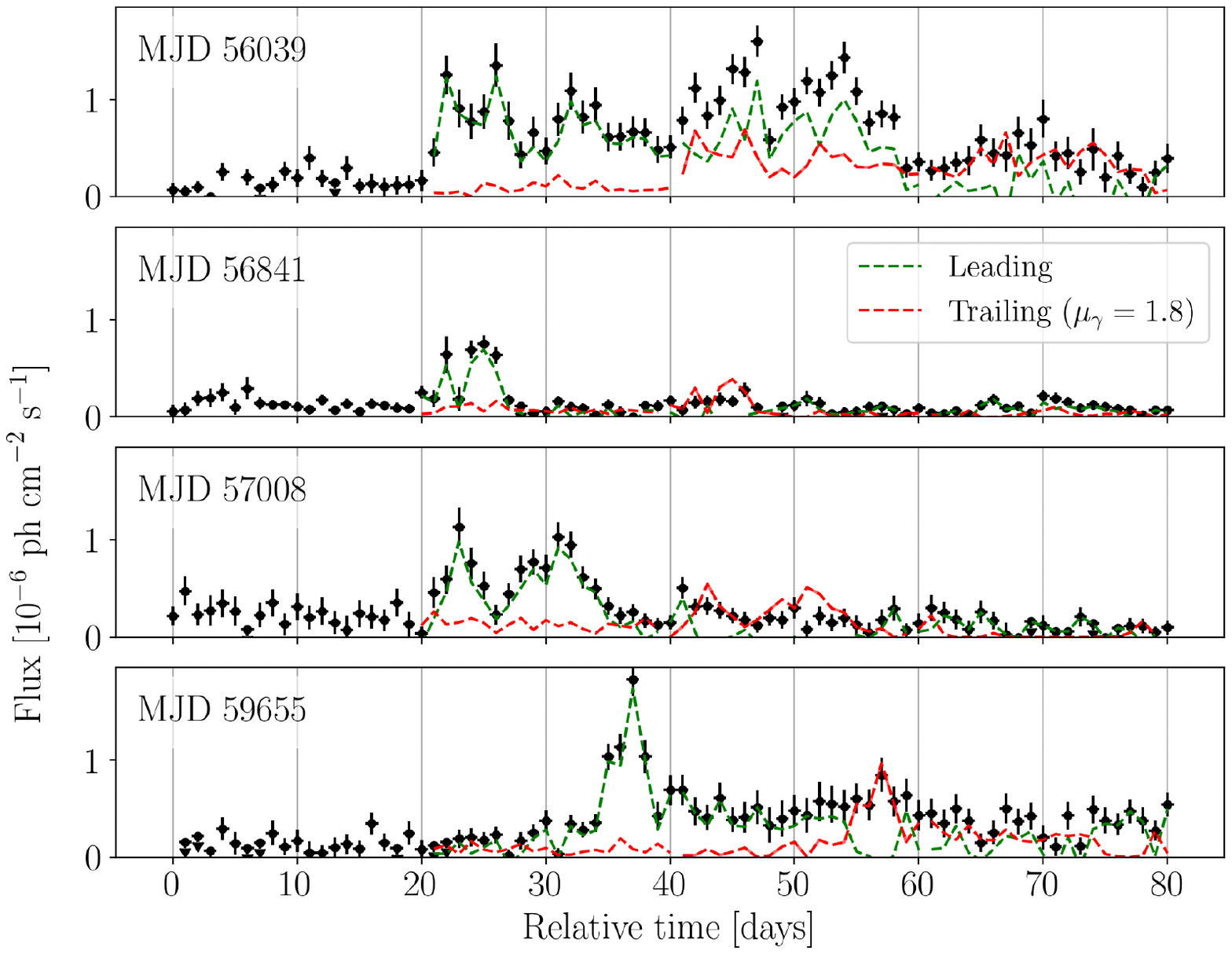}
    \caption{
    For four relatively  sharp, individual flares, the original LC (black points) is decomposed in the leading (green) and trailing (red) components. The latter is obtained by shifting the leading one of \tab, and assuming a \magRatioG=1.8 (Section \ref{sec:micro}). Each panel reports in the top left corner the segment’s starting MJD.
    }
    \label{fig:micro_macro_lensing}
\end{figure*}
     
\section{Comparison with previous results}
\label{sec:comparison}

Previous analysis of \Fermi-LAT  data yielded inconsistent estimates of the time delay \tab, with values ranging between $\sim19$ and $\sim27$ days  \citep{Barnacka11,Barnacka:2015,Neronov15,Abhir:2021,Agarwal:2025}, along with possible detections of microlensing effects. 
Most of them focused on early mission data up to $\sim$~MJD~57000, that corresponds to data up to our mid-segment 3, and did not account for limitations such as the Sun's \g-ray emission and potential impact of data gaps. These limitations also applied to our initial study, based on the time interval MJD $54700 - 55750$, where we found no clear trailing component in the \g-ray LC (\paperI). 

We have now re-analyzed the same interval using the new method outlined in Section~\ref{sec:time_series} and find a time patter consistent with \tab$\sim20$~days.
As explained in the following, the differing results can be reconciled when considering the improved signal-to-noise time-series data used in this work. We adopt the latest LAT IRFs, \texttt{Pass8}, instead of the older \texttt{Pass6}, apply LC detrending, use a stricter zenith angle cut ($90^{\circ}$ versus $105^{\circ}$). In addition, we show that times affected by solar contamination within the ROI and data gaps can make it challenging to detect and interpret lensing effects in the blazar.

\subsection{Confirmation of \tab$\sim20$~days}
Figure~\ref{fig:paperI} (right) shows two results: the DACF of the segment analysis (MJD $54700 - 55750$), replicating our original study with no censoring of data (light blue), and the analysis after removing the times when the Sun is near the ROI (blue). The peak at about 20 days, with local significance $\gtrsim3\sigma$, is consistent with the findings presented in this work. In fact, a similar peak at $\sim$20 days was already visible in the DACF computed from a 150-day LC binned at 12 hours, as shown in Figure~3 of \paperI. In this present work, we deliberated excluded this data segment initially to avoid potential biases, using it only later as a consistency check. 


\subsection{Contamination by solar activity}
\label{sec:sun_contamination}
An elevated flux state was observed around MJD~55560, previously attributed solely to intrinsic blazar activity. However, our re-analysis reveals substantial solar activity observable during this 20-days period, see Figure \ref{fig:paperI} (top). If not properly accounted for, solar contamination can distort the blazar flux measurements and bias estimates of the magnification ratio. Our findings are robust in this regard, as we conservatively excluded data potentially affected by solar emission.

\subsection{Reduced \Fermi-LAT exposure may hamper detection of leading flare component}
\label{sec:crabToO}
 
\begin{figure*}[htb]
\centering
\setlength\tabcolsep{3pt}
\adjustboxset{width=\linewidth, valign=m, margin=0pt 3pt 0pt 3pt}
\begin{tabularx}{1.05\linewidth}{@{} X @{}}
\adjustimage{}{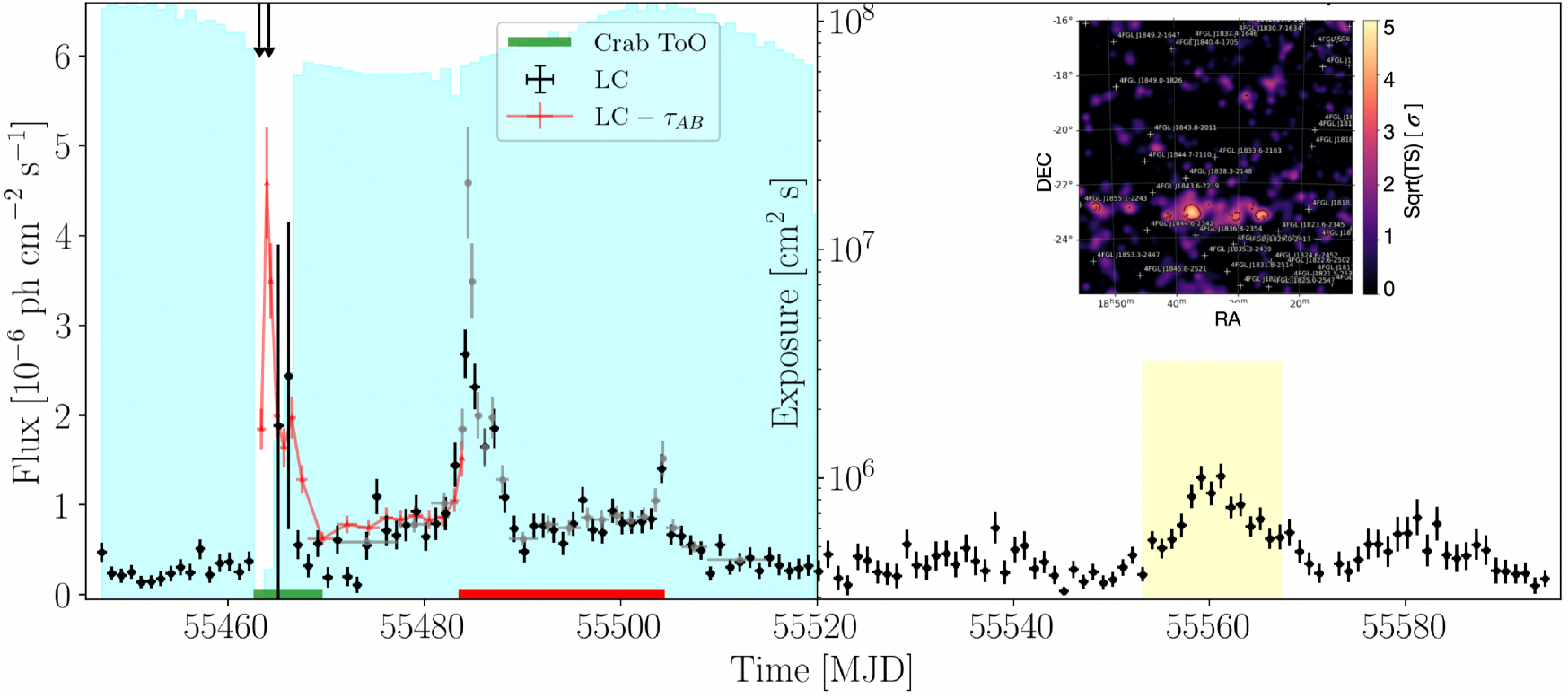} \\
\end{tabularx}
\begin{tabularx}{\linewidth}{@{} XX @{}}
 \parbox{0.95\linewidth}{\caption{Top: A leading-flare component may have been undetected due to low exposure (cyan) toward the region of \pks, during a Crab ToO observation (green, Section \ref{sec:crabToO}).      
    The elevated $\sim$-20-day flux state observed around MJD~55560, coincides with the passage of the Sun close ($<2^\circ$) from the blazar (yellow area, Sec.~\ref{sec:sun_contamination}). Inset: when inspecting the LAT test-statistics map of the ROI integrating data over the time interval highlighted in yellow, unmodeled solar emission is visible at declination $\sim -22.83^{\circ}$. 
    Right: DACF of the LC interval MJD~$54700-55750$ studied in \paperI\ (see Section \ref{sec:comparison}) with time intervals during which the Sun is within $8^{\circ}$ of the ROI center excluded from the analysis (light blue), and included (blue)}
    \label{fig:paperI}
    } 
    & \adjustimage{width=1.15\linewidth}{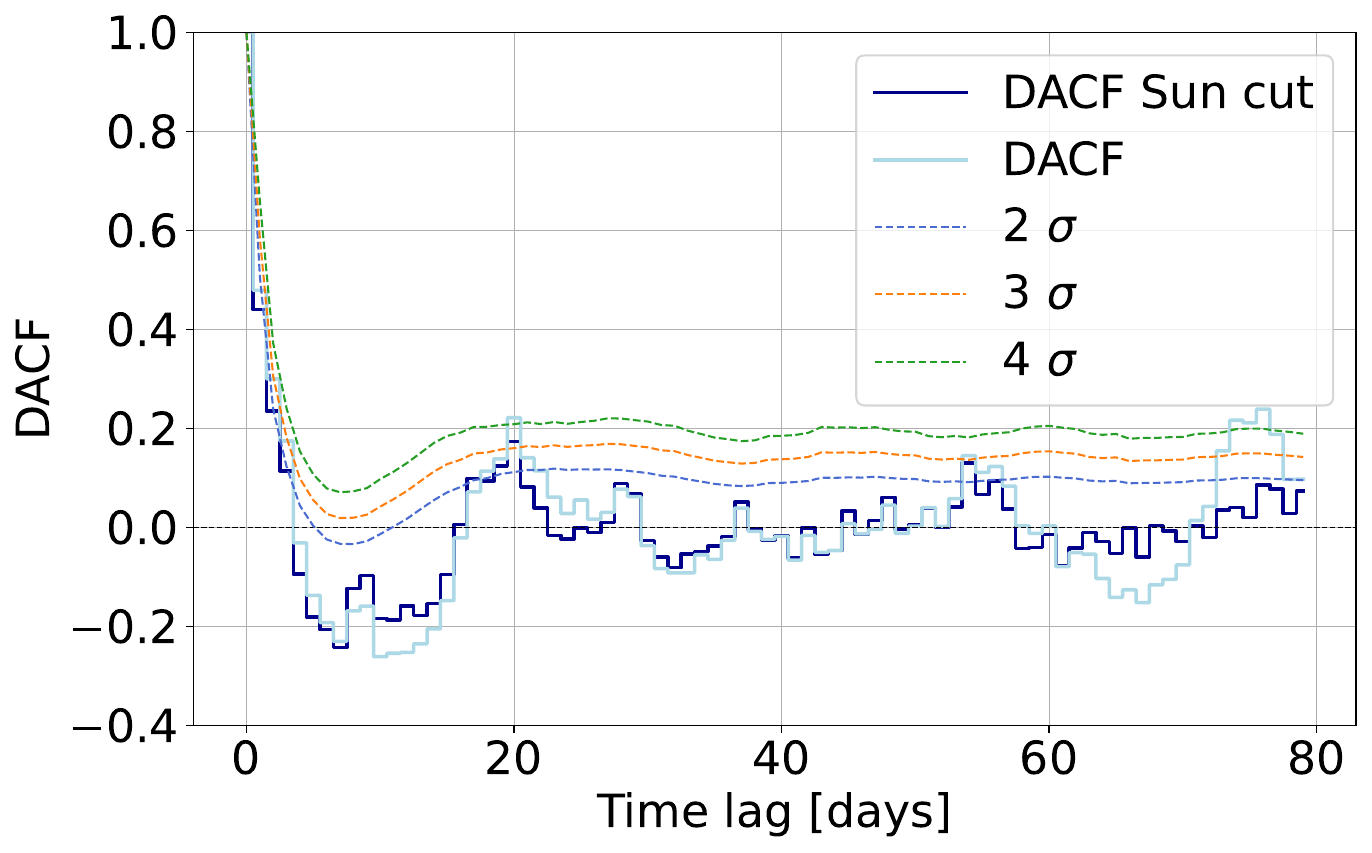} 
\end{tabularx}
\end{figure*}

A bright flare around MJD~55485 lacks a corresponding lensed counterpart (Figure~\ref{fig:paperI} top). Our re-analysis suggests it may be the delayed component of a flare that went undetected due to reduced LAT exposure 20 days earlier during a Crab Nebula ToO (MJD 55462.67–55469.66, green horizontal line). When shifting the LC (gray, adaptive binning) by $-$\tab~($-20.26$ days) assuming \magRatioG=1.0, the putative leading component (red) remains consistent with LAT upper limits. At that time, the only other \g-ray instrument in operation that could had detected the flare, the AGILE-Gamma-Ray Imaging Detector, was operating in spinning mode, observing only a fraction of the sky \citep{Agile_atel:2010}.


\section{Microlensing and \magRatioG\ variability}
\label{sec:micro}
Earlier works identified variability in the \g-ray magnification ratio (\magRatioG$\sim2-7$) in \pks, which was interpreted as microlensing by stars in the lensing galaxy affecting compact emission regions \citep{Neronov15,Barnacka:2015,Agarwal:2025}. However, this conclusion relied on focusing on a few isolated flare episodes and a simplified flare-based flux ratio method to infer the \magRatioG\ between the leading and trailing lensed components. In light of our findings and estimates of \tab$\sim20$ days and \magRatioG$~\lesssim~1.8$, they can be reconciled with macrolensing effect.
The novel method introduced in Section \ref{sec:macrolensing} provides a more objective approach to quantify \magRatioG, based on the long-term LC information. Upon qualitatively inspecting four LC segments previously reported to exhibit microlensing or \magRatioG\ variability, we do not find convincing indications of such effects. The corresponding 1-day bin LC segments are plotted in black in Figure~\ref{fig:micro_macro_lensing}. For each 20‑bin interval, we decompose the trailing component (red) assuming a fixed flux ratio\footnote{
    If a point-like source is split into multiple images, each image (e.g., A and B) is magnified by a different magnification factor relative to the flux of the unlensed source. The ratios of these magnification factors correspond to the flux ratios of the images \citep{Schneider:1992}.
}
of the leading/trailing components,  \magRatioG\,=\,1.8.
The leading component (green) is computed as the difference between the observed flux and the trailing component. The resulting decomposition traces closely the overall LC trend within statistical measurement uncertainties, and supports lower macrolensing magnification ratios, in agreement with radio-based estimates. 

In addition, we note the following. The first flare considered in \citet{Neronov15}, at about MJD~55485, was treated in that work as a leading component, but it may be a trailing one, at least in part. Its actual leading component would fall during a period of very low LAT exposure due to a Crab ToO (Section~\ref{sec:crabToO}).
The second flaring epoch considered, at about MJD~55561, occurred when the Sun was close to the blazar, possibly contaminating blazar's flux measurements (Section~\ref{sec:sun_contamination}, Figure~\ref{fig:paperI}).
Finally, for the third flare at about MJD~56867, the LC is consistent with a trailing component having a \tab~$\sim$20 days and \magRatioG$~\lesssim~1.8$, as found here (Figure \ref{fig:micro_macro_lensing}).

\section{Discussion and Conclusions}
\label{Sec:conclusions}

We presented an analysis of $\sim 13$-yr of LAT data, with improved time-series techniques, of the gravitationally lensed system \pks. We shed light on the macrolensing properties of this system, reconciling previous controversial findings at \g-rays. The main results are:
\begin{itemize}
\item 
Applying the time-series analysis to six, non-overlapping data segments consistently reveals a recurrent pattern in the data, with best-fit estimate \tab~$= 20.26 \pm 0.62$. The likelihood of this occurring by chance is as low as $2.5 \times 10^{-5}$. This lag in the LC, attributed to gravitational lensing, corresponds to the delay in arrival times between the \g-ray emission from images A and B. Based on simulations, there is no evidence for variations of this lag at \g-ray energies. 

\item 
Additional delay patterns emerge from the time series analysis, but none is statistically significant or convincingly linked to a possible third lensed image. Such a third image would be expected to be much fainter than the two main ones \citep{Muller:2020}.

\item In the DACFs of the blazar, we observe a characteristic triplet of peaks. Simulations reveal that this pattern naturally arises by the interplay between rapid stochastic variability and a time-lag component. This corroborates the genuineness of the time-lag observed in \pks. It also offers a new approach to identify and study time-patterns in LCs.

\item 
Our novel approach to estimate \magRatioG\ exploits the full LC information by using the flux-normalized LC segments and statistical modeling of DACF peak heights to reduce bias and avoid misinterpretation due to isolated flares or incomplete sampling. The estimated \magRatioG~$\lesssim 1.8$, is consistent with radio-based estimates. In addition, we find no supporting evidence for variability in \magRatioG\ or microlensing; we discuss that earlier different findings may be reconciled with specific analysis choices.
\item
The \tab\ value inferred at \g-rays is in tension ($\sim2 \sigma$) with the average value from radio estimates, as well with predictions from lens modeling, which are based on radio observations too \citep{Muller:2020}. If the \g-ray/radio delays difference arises from an intrinsic offset between their respective emission regions, then, within a singular isothermal sphere lens model, a $\sim10$\% difference in the time delay implies a $\sim100$ pc offset (projected, in the source plane). 
\end{itemize}

For \pks, the \g‑ray/radio emission region offset is comparable to that reported for the only other known \g‑ray–lensed blazar,  B$0218+357$ \citep[$\sim70$ pc;][]{Cheung:2014}. However, a later reanalysis of the radio data shows that, for B$0218+357$, the radio and \g-ray time delays are statistically consistent \citep{Biggs:2018}. Such separations remain unusually large compared to typical parsec-scale offsets found in other blazars \citep[][]{Kramarenko:2022}. 
%
%
A time delay difference between radio and \g-rays does not necessarily imply different emission regions. 
In blazars, it is often observed that \g-rays originate upstream while the accompanying radio emission emerges further away from the black hole. Due to these opacity effects, radio variations often lag behind \g-ray ones, especially in FSRQs, introducing an intrinsic observational offset.

Our study demonstrates that long-term, well-sample LCs, coupled with time-series analysis, enable competitive constraints on gravitational lensing parameters, providing a unique probe into the structure of blazar jet emission regions on otherwise inaccessible spatial scales. 
These systems hold potential for time-delay cosmography \citep{Birrer:2024}. With improved lens models and more precise estimate of the location of the emission region sites, they could provide independent determinations of the Hubble constant, offering complementary insights to those from local distance ladders.

\begin{acknowledgments}
This work was supported by the European Research Council, ERC Starting grant \texttt{MessMapp}, S.B. Principal Investigator, under contract no. 949555, and partially by the German Science Foundation (DFG FOR 5195, grant No. 443220636).
%
%
The \textit{Fermi} LAT Collaboration acknowledges generous ongoing support from a number of agencies and institutes that have supported both the development and the operation of the LAT as well as scientific data analysis. These include the National Aeronautics and Space Administration and the Department of Energy in the United States, the Commissariat \`a l'Energie Atomique and the Centre National de la Recherche Scientifique / Institut National de Physique Nucl\'eaire et de Physique des Particules in France, the Agenzia Spaziale Italiana and the Istituto Nazionale di Fisica Nucleare in Italy, the Ministry of Education, Culture, Sports, Science and Technology (MEXT), High Energy Accelerator Research Organization (KEK) and Japan Aerospace Exploration Agency (JAXA) in Japan, and the K.~A.~Wallenberg Foundation, the Swedish Research Council and the Swedish National Space Board in Sweden. Additional support for science analysis during the operations phase is gratefully acknowledged from the Istituto Nazionale di Astrofisica in Italy and the Centre National d'\'Etudes Spatiales in France. This work performed in part under DOE Contract DE-AC02-76SF00515.  
\end{acknowledgments}

\begin{contribution}
The three corresponding authors contributed substantially and equally to the analysis, interpretation and preparation of the manuscript.
\end{contribution}

%
\facilities{Fermi(LAT)}

\software{  
          fermipy \citep{Wood:2017}, astropy \citep{Astropy:2013}, TopCat \citep{Taylor:2005}
          }


\appendix

\begin{figure*}[h]
\centering
\begin{minipage}{0.46\textwidth}
    \captionof{figure}{\small
    \textbf{Illustration of the triplet effect} (see Section \ref{sec:triplet}).
    \textbf{Top:} In the presence of a time lag, two distinct flares (a, c) have corresponding trailing components (b, d) at the lag time. These generate overlapping variability, which increases the stochastic contribution to the LC and the DACF.
    \textbf{Middle:} 
    In the DACF, cross-talk between these flares arises. The red-arrow peak represents the genuine lag, while peaks 1, 2, and 3 emerge from the interplay between noise and this ``genuine'' lag, forming a triplet structure.
    \textbf{Bottom:} 
    This  triplet effect manifests in the presence of sufficiently pronounced flares, as clearly visible in segment 4 of \pks, where three distinct peaks occur separated by the characteristic time lag. Their actual amplitude depends on the magnification ratio, and are further influenced by overlapping variability in a short LC.
    }
        \includegraphics[width=\linewidth]{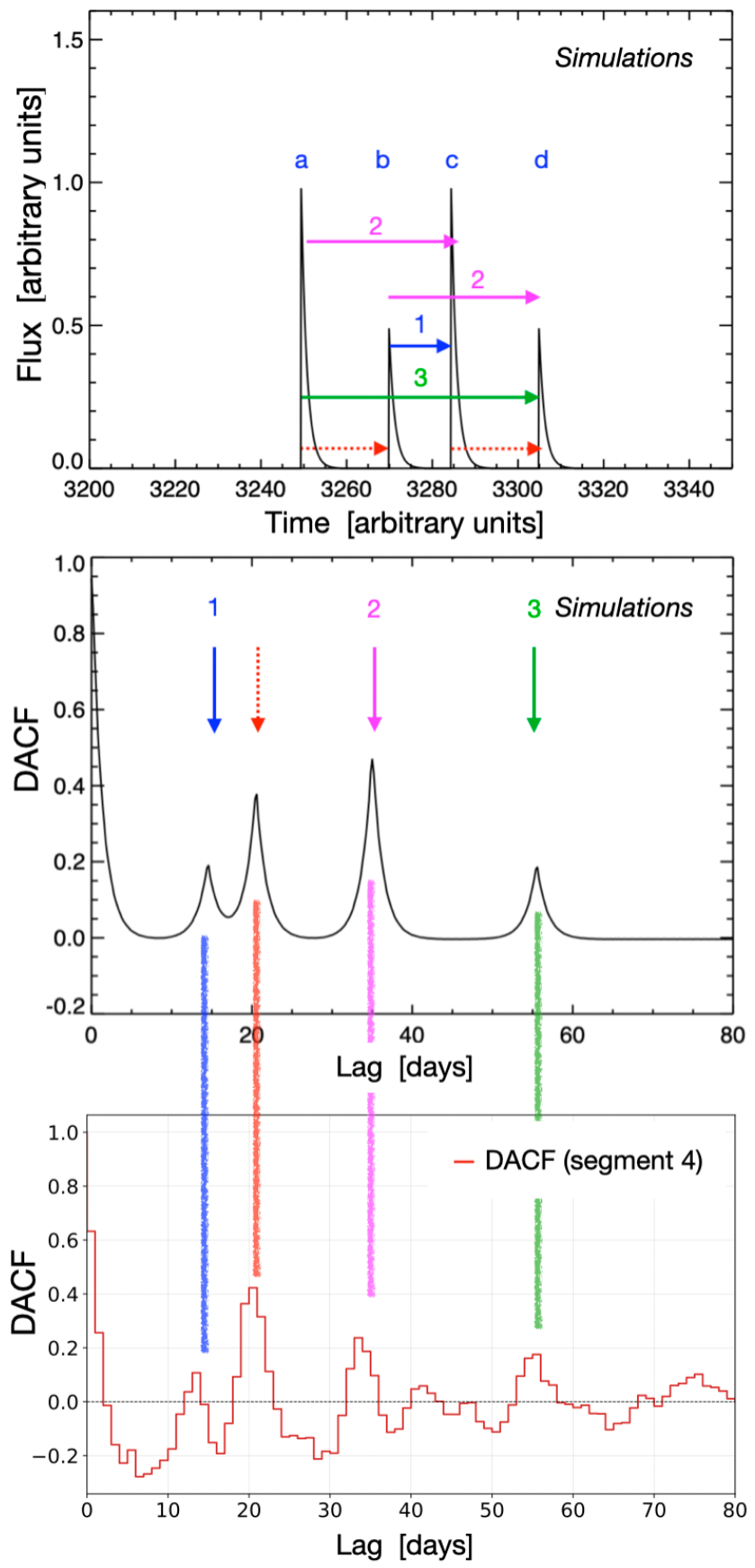}
    \label{fig:triplet}
\end{minipage}
\hfill
\begin{minipage}{0.46\textwidth}
\captionof{figure}{\small
    \textbf{
DACF triplets in simulations and in the LC of \pks} (see Section \ref{sec:triplet}). 
    \textbf{Top:} 
    Distribution of triplet separations in simulated LCs with (blue) and without (pink) a delayed component.
    \textbf{Middle:} 
    Distribution of triplet separations in the mixed correlations between the 6 active epochs in the LC of \pks. 
    \textbf{Bottom:} 
     Average DACF of the mixed segment correlations used to obtain the distribution in the middle plot. 
}
    \includegraphics[width=.99\linewidth]{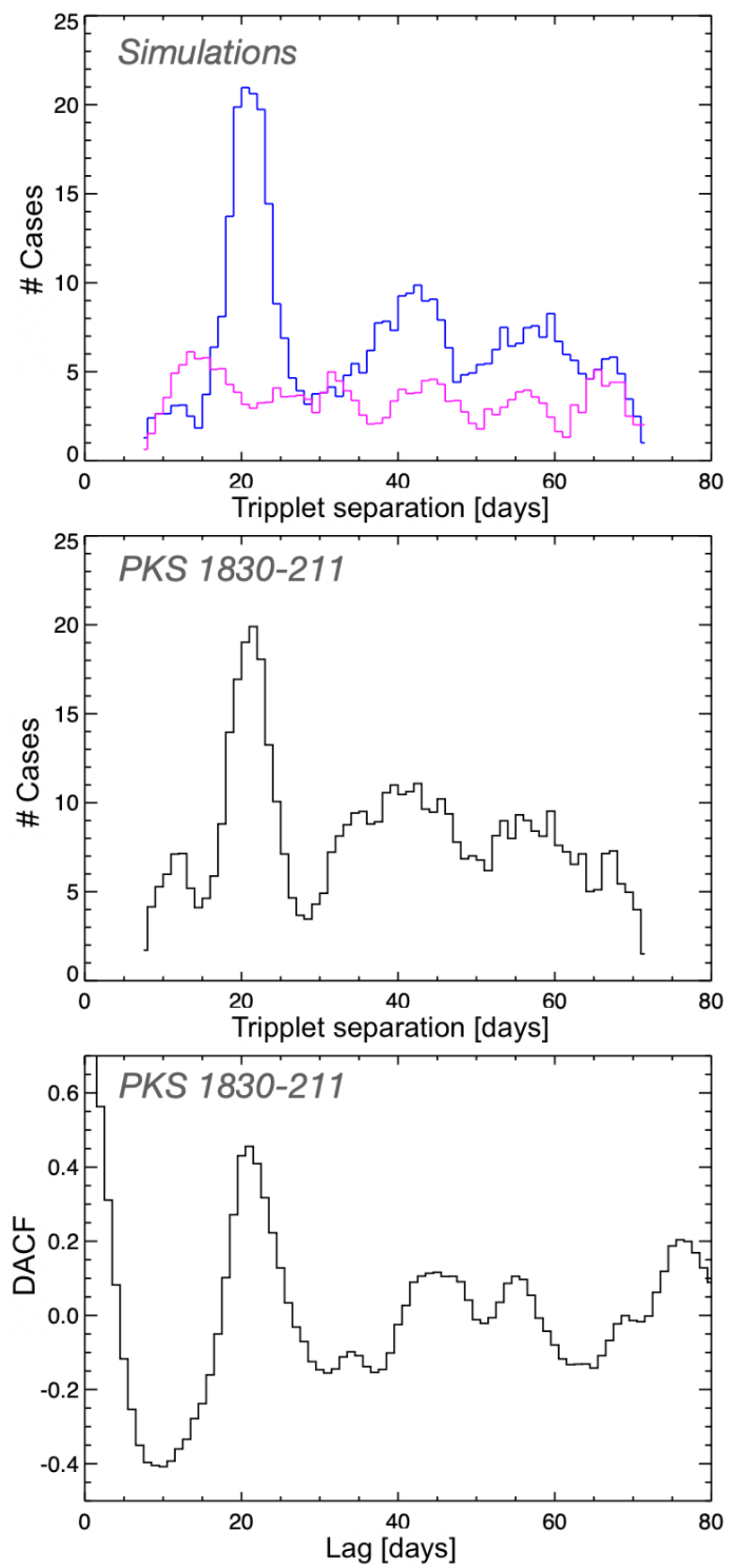}
\label{fig:mixedccfs}
\end{minipage}%
\end{figure*}

\section{Benchmarking against comparable bright \g-ray blazars}
\label{app:checks}

The same methodology used for \pks\ was applied to two other bright variable \g-ray blazars,  CTA~102 (4FGL~J2232.6$+$1143) and 3C~454.3 (4FGL~J2253.9$+$1609), using segments matching the duration of those from \pks. This allows us to assess potential systematics in the analysis, such as the impact of adopting a fixed PSD spectral index in the simulations. As described in section \ref{Subsec:time_segments}, the large range in flux level over the LC also means that the overall DACF is heavily dominated by the brightest epoch, which is the flaring episode around MJD~58600. While using a fixed spectral index, derived from the whole LC segment considered, for the simulations is a reasonable assumption at first order, it does not account for possible variations of the spectral index within time sub-intervals.  Applying our methodology on other blazars assesses whether DACF features, similar to those observed in \pks, can arise from suboptimal PSD modeling in the mock LCs.

\begin{figure}
    \centering
    \includegraphics[width=0.95\linewidth]{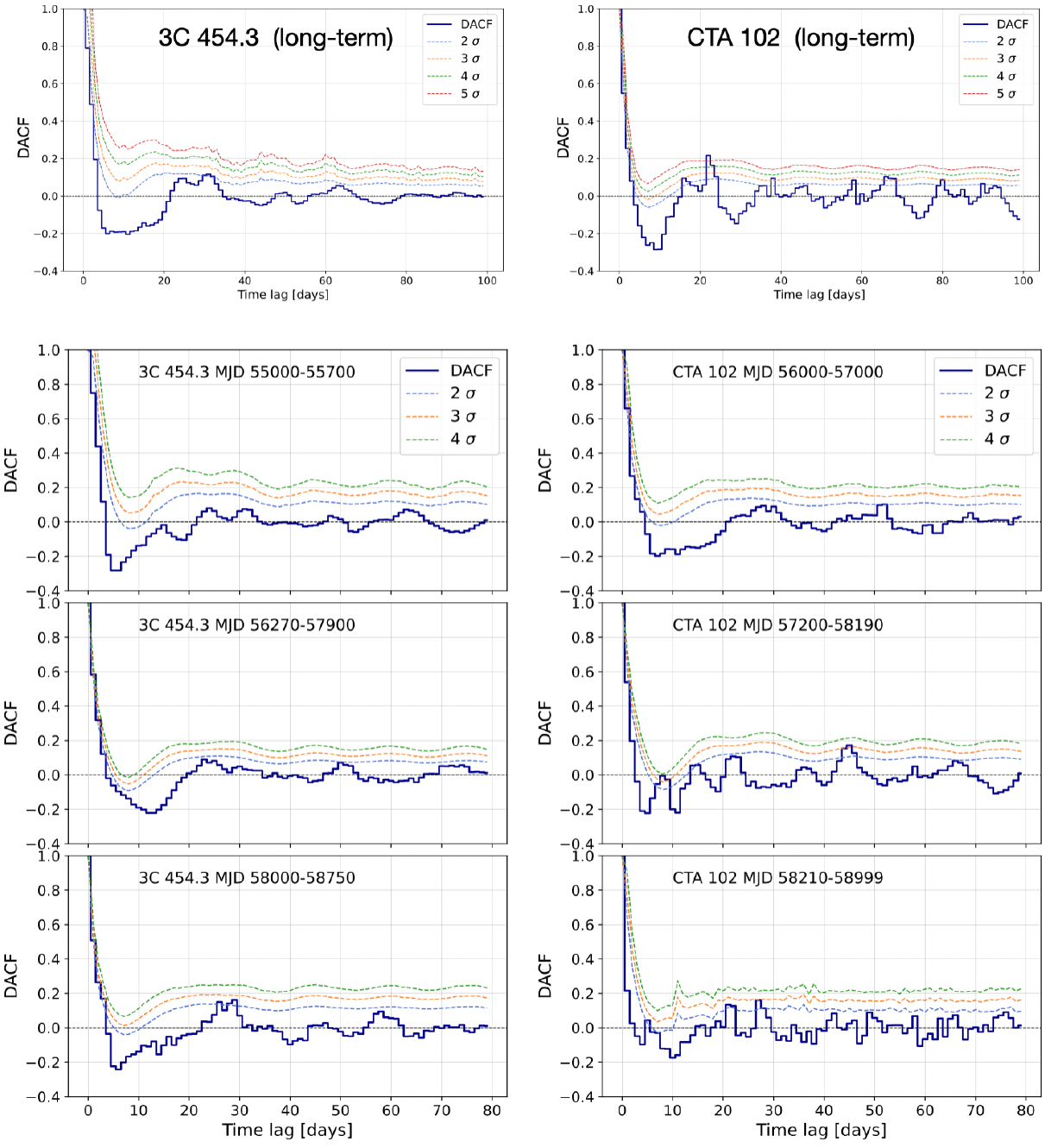}
    \caption{
    Top: DACFs of the long-term LCs (MJD 54682–60178) for the two brightest LAT \g-ray blazars, 3C~454.3 (left) and CTA~102 (right).
    Bottom: DACFs and significance curves (Section \ref{subsec:local_sign_method}) from three LC segments of 3C~454.3 (left; MJD 55000–55700, 56270–57900, 58000–58750) and CTA~102 (right; MJD 56000–57000, 57200–58190, 58210–58999).
    }
    \label{fig:full_other}
\end{figure}

In the case of 3C~454.3, no time lag stands out  either from the long-term LC or the individual segments DACFs (left Figure~\ref{fig:full_other}). On the other hand, the DACF of the long-term CTA~102 LC displays several time lags exceeding $2\sigma$ and up to above $5\sigma$ (right Figure~\ref{fig:full_other}). Inspecting the LC (not shown), these are likely due to stochastic variability, as they coincide with the separation between major active states. As a reprove, none of these lags appear consistently when analysing segments of active phases as done for \pks. As noted in Sec.~\ref{subsec:global_sign}, random correlations from stochastic processes can produce peaks at arbitrary time lags, whereas a genuine delayed component would show as a recurrent peak at the same lag across segments. In both objects, no such repeated peaks above $2\sigma$ significance are found. This supports the reliability of the adopted methodology and results.

\bibliography{sample701}{}
\bibliographystyle{aasjournalv7}

\end{document}